# Improved adaptive sparse channel estimation using mixed square/fourth error criterion

Guan Gui*, Li Xu, and Shinya Matsushita

Department of Electronics and Information Systems, Akita Prefectural University, Akita, 015-0055 Japan

* Corresponding author. Email: guitohoku2009@gmail.com

**Abstract**

Sparse channel estimation problem is one of challenge technical issues in stable broadband wireless communications. Based on square error criterion (SEC), adaptive sparse channel estimation (ASCE) methods, e.g., zero-attracting least mean square error (ZA-LMS) algorithm and reweighted ZA-LMS (RZA-LMS) algorithm, have been proposed to mitigate noise interferences as well as to exploit the inherent channel sparsity. However, the conventional SEC-ASCE methods are vulnerable to 1) random scaling of input training signal; and 2) imbalance between convergence speed and steady state mean square error (MSE) performance due to fixed step-size of gradient descend method. In this paper, a mixed square/fourth error criterion (SFEC) based improved ASCE methods are proposed to avoid aforementioned shortcomings. Specifically, the improved SFEC-ASCE methods are realized with zero-attracting least mean square/fourth error (ZA-LMS/F) algorithm and reweighted ZA-LMS/F (RZA-LMS/F) algorithm, respectively. Firstly, regularization parameters of the SFEC-ASCE methods are selected by means of Monte-Carlo simulations. Secondly, lower bounds of the SFEC-ASCE methods are derived and analyzed. Finally, simulation results are given to show that the proposed SFEC-ASCE methods achieve better estimation performance than the conventional SEC-ASCE methods.

## 1. Introduction

Accurate channel state information (CSI) of frequency-selective fading channel is required for broadband signal transmission, which is one of the mainstream techniques in the next generation communication systems [1][2]. A promising approach to obtain accurate CSI is an adaptive channel estimation (ACE) using square error criterion (SEC) based standard least mean square error (LMS) algorithm, whose general structure is shown in Fig. 1 [3]. The advantage of the LMS is its low complexity and easy implementation. However, the steady-state mean square error (MSE) performance is greatly affected by random scaling of input training signal, signal transmit power and noise power [3]. One main reason is the use of invariable step-size (ISS) which may not balance efficiently in these variables. Therefore, to improve the steady-state MSE performance, a fourth error criterion (FEC) based least mean fourth error (LMF) filtering algorithm with variable step-size (VSS) was proposed in [4]. Indeed, the initial step-size should be chosen very small for LMF to ensure the stability of gradient descend even if VSS adjusts its convergence speed with updating MSE performance. Hence, FEC-based LMF algorithm still has high computational complexity [4].

To complement the shortcomings of LMS algorithm and LMF algorithm, a mixed square/fourth error criterion (SFEC) based least mean square/fourth error (LMS/F) algorithm was first proposed in [5] and further developed in [6]. Comparing to LMS algorithm, LMS/F algorithm improves the steady-state MSE performance and achieves stability property without sacrificing the simplicity of LMS algorithm. However, these developed algorithms (i.e., LMS, LMF and LMS/F) focus on different error criterions but without considering practical prior channel structures which could be utilized to further improve estimation performance.

Recently, a number of channel measurements have verified that broadband channels often exhibit sparse structure [2], [7], [8]. Consider an example with signal bandwidth of 7.56 MHz with its carrier frequency at 770 MHz, which is shown in Fig. 2. The six-tap Vehicular B channel model is adopted with the maximum delay spread of 20 $\mu s$ [2]. According to Nyquist-Shannon sampling theorem, discrete channel length is equal to 30. Hence, in this kind of broadband channels, a few taps have nonzero coefficients and most of them have zeroes. Traditional ACE methods with either LMS or standard LMS/F algorithm focus on noise interferences as well as gradient error but without considering channel structure. To further improve the estimation performance, one of effective approaches is to develop adaptive sparse channel estimation (ASCE) methods so that they can exploit the inherent channel sparsity as well as can realize similar functions of traditional ACE methods. In other words, the further performance gain of channel estimation could be obtained by taking advantage of a prior structure information of channels.

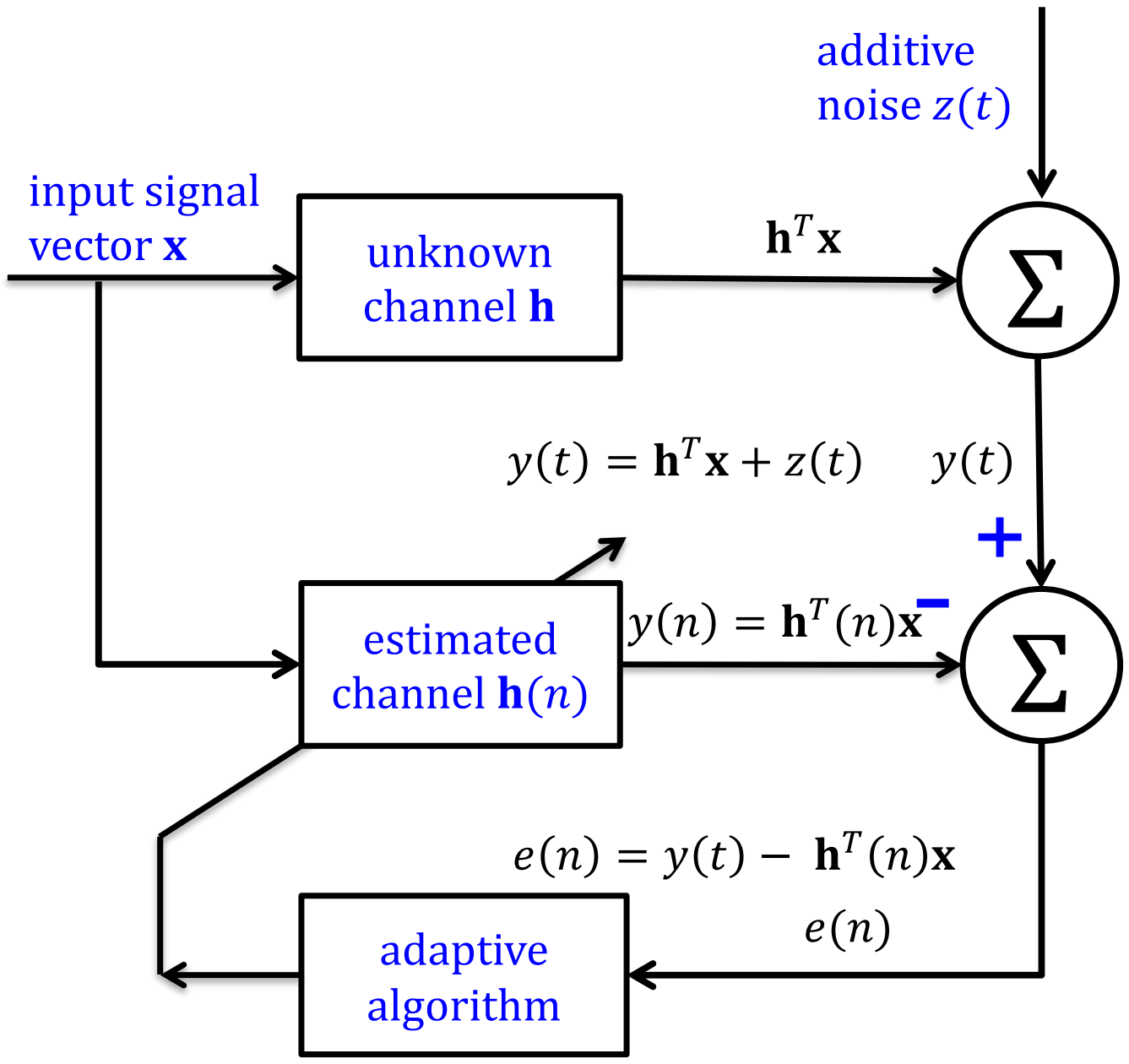

Fig. 1. ACE for estimating unknown channels.

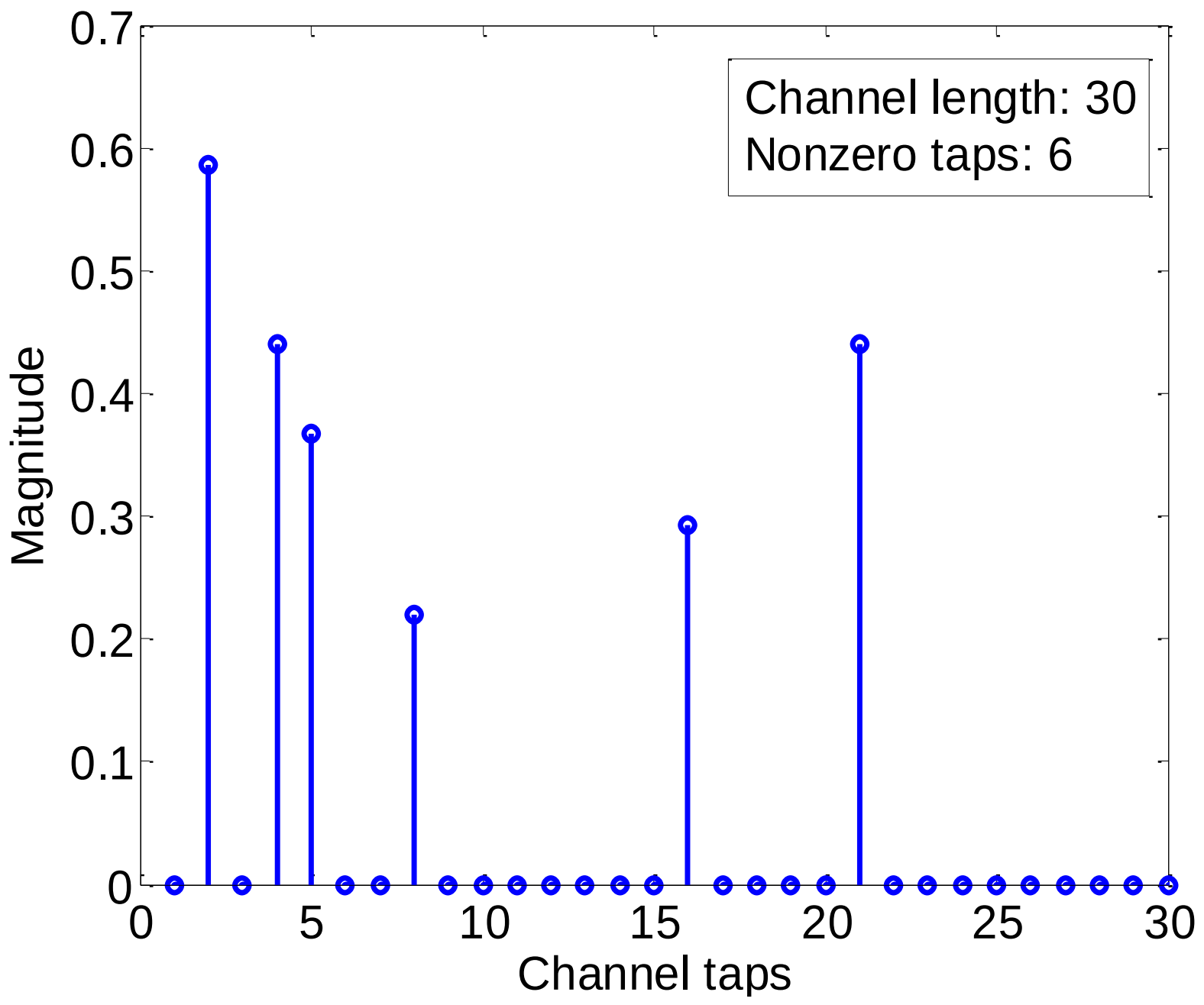

Fig. 2. A typical example of sparse multipath channel.

To exploit the channel sparsity, $\ell_p$-norm-penalized sparse LMS/F filtering algorithm (LP-LMS/F) was proposed in [9]. In [9], we conduct the performance confirmation of LP-LMS/F by means of computer simulations. However, mathematical analysis for LP-LMS/F algorithm is very challenging due to the fact that $\ell_p$-norm is nonconvex function [10]. Inspired by least absolute shrinkage and selection operator (LASSO) algorithm [11], an adaptive sparse channel estimation (ASCE) method is first proposed with sparse LMS/F algorithms by introducing an $\ell_1$-norm penalized constraint function, namely zero-attracting least mean square/fourth filtering (ZA-LMS/F) algorithm. Then, based on the recent development of the reweighted $\ell_1$-norm sparse constraint function [12], a reweighted ZA-LMS/F (RZA-LMS/F) algorithm is developed as well. Note that a part of the work was presented in [13] but without considering theoretical analysis. In this paper, initial work in [13] is extended and sparse LMS/F filtering algorithms are studied comprehensively from computer simulations and theoretical analysis. Different to [9], a profound mathematical analysis for sparse LMS/F algorithms are given by adopting $\ell_1$-norm based convex sparse constraint functions, ZA and RZA. The main contributions of this paper are summarized as follows:

1) A derivation of the steady-state mean square error (MSE) of standard LMS/F algorithm.

2) A construction of the sparse penalized cost functions of sparse LMS/F algorithms to obtain sparse solutions.
3) A proposal of ZA-LMS/F and RZA-LMS/F.
4) An adaptive selection of VSS for sparse LMS/F algorithms according to estimation error and given threshold parameter.
5) A development of Monte Carlo-based selection of the approximate optimal regularization parameter for exploiting channel sparsity.
6) Comprehensive computer simulation to validate the effectiveness of the proposed algorithms. The evaluation of the impacts of channel sparsity, K, and reweighted factors of RZA-LMS/F on the steady-state MSE performance.

The remainder of the rest paper is organized as follows. A system model is described and standard LMS/F is pointed out and the drawback of the LMS/F is uncovered via theoretical analysis in Section 2. In section 3, sparse ASCE using ZA-LMS/F is proposed and improved ASCE using RZA-LMS/F is highlighted as well. Computer simulations are presented in Section 4 in order to evaluate and compare performances of the proposed ASCE methods. Finally, Section 5 concludes the paper.

**Notations**: Capital bold letters and small bold letters denote matrices and row/column vectors, respectively; $(\cdot)^{\mathrm{T}}$, $(\cdot)^{\mathrm{H}}$, $(\cdot)^{-1}$and $|\cdot|$ denote the transpose, conjugate transpose, matrix inversion and absolute operations, respectively; $E[\cdot]$ and $E[\cdot|\cdot]$ denote the expectation and conditional expectation, respectively; assume any vector $\mathbf{h}=[h_0,\ldots,h_l,\ldots,h_{N-1}]^T$, $\|\mathbf{h}\|_p$ denotes $\ell_p$ norm, i.e., $\|\mathbf{h}\|_p=(\sum_l |h_l|^p)^{1/p}$, $p=\{1,2\}$ ; $\mathrm{sgn}(\mathbf{h})$ is a component-wise function which is defined as $\mathrm{sgn}(h_l)=1$ for $h_l>0$, $\mathrm{sgn}(h_l)=0$ for $h_l=0$, and $\mathrm{sgn}(h_l)=-1$ for $h_l<0$; and $\tilde{\mathbf{h}}$ represents the channel estimate of $\mathbf{h}$.

## 2. Standard LMS/F Algorithm and Its Drawbacks

### *2.1. Standard LMS/F Algorithm*

The standard LMS/F algorithm [5] is briefly revisited. Consider a baseband frequency-selective fading wireless communication system. The finite impulse response (FIR) sparse channel vector $\mathbf{h}$ is N-length and it is supported only by $K$ nonzero channel taps. Assume that an input training signal $\mathbf{x}$ is used to probe the unknown sparse channel at the time t. At the receiver side, observed signal $y(t)$ is given by

$$y(t)=\mathbf{h}^T\mathbf{x}+z(t), \tag{1}$$

where $\mathbf{x}=[x(t),x(t-1),...,x(t-N+1)]^T$ denotes the vector of input signal $x(t)$ with $\mathcal{CN}(0,\sigma_x^2)$ and $z(t)$ is the additive white Gaussian noise (AWGN) variable satisfying $\mathcal{CN}(0,\sigma_n^2)$, which is assumed to be mutually independent with the input training signal $x(t)$. The objective of ASCE is to adaptively estimate the unknown sparse channel vector $\mathbf{h}$ using training signal vector $\mathbf{x}$ and observed signal $y(t)$. By defining the estimation error at the n-th update step by $e(n)$, the cost function of the standard LMS/F is given as [6]

$$G_{lmsf}(n)=\frac{1}{2}e^2(n)-\frac{1}{2}\lambda\ln\left(e^2(n)+\lambda\right), \tag{2}$$

where $\lambda$ is a positive threshold parameter which controls the convergence speed and stability of the LMS/F. The low convergence speed means high stability and vice versa. From Eq. (2), the corresponding update equation of LMS/F is given by

$$\tilde{\mathbf{h}}(n+1)=\tilde{\mathbf{h}}(n)+\mu\frac{\partial G_{lmsf}(n)}{\partial\tilde{\mathbf{h}}(n)}=\tilde{\mathbf{h}}(n)+\frac{\mu e^3(n)\mathbf{x}}{e^2(n)+\lambda}, \tag{3}$$

where $\mu$ is the initial update step-size. Different from the LMS [3], one can find that the step-size of LMS/F depends on the estimation error $e(n)$ and given parameter $\lambda$ as follows:

$$\mu(n)=\frac{\mu e^2(n)}{e^2(n)+\lambda}. \tag{4}$$

By pugging Eq. (4) into Eq. (3), the update equation of LMS/F can be rewritten as

$$\tilde{\mathbf{h}}(n+1)=\tilde{\mathbf{h}}(n)+\mu(n)e(n)\mathbf{x}, \tag{5}$$

The step size $\mu(n)$ balances the steady-state MSE performance and the convergence speed of the gradient descend method of LMS/F.

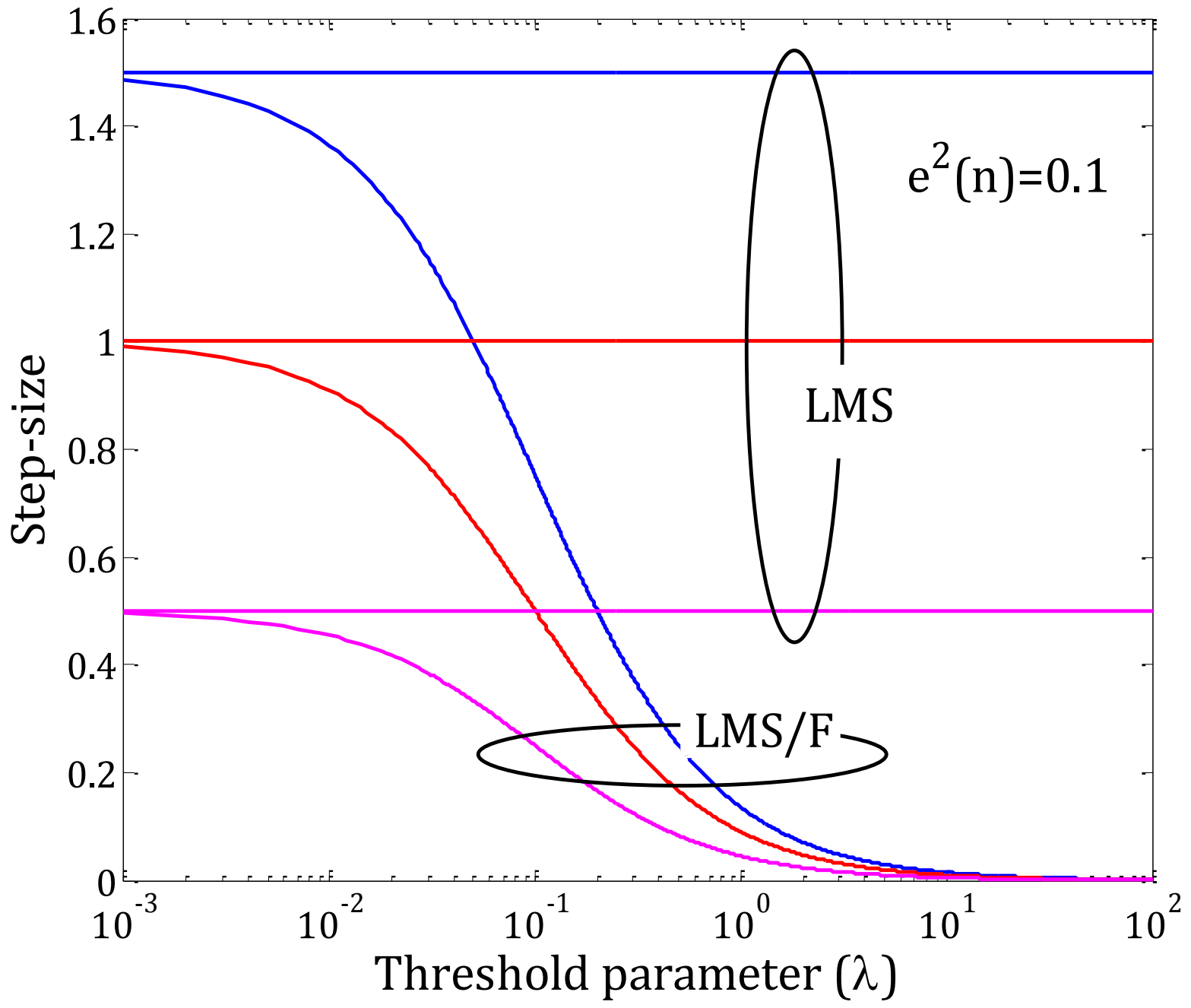

Fig. 3. Threshold parameter ($\lambda$) controls the variable step-size of LMS/F algorithm.

### *2.2. Drawbacks of the LMS/F Algorithm*

The VSS in Eq. (4) indicates that the LMS/F behaves like the standard LMF with a step size of $\mu/\lambda$ for $\lambda \sim e^2(n)$ and it degenerates to the standard LMS algorithm with a step size of $\mu$ for $\lambda \sim e^2(n)$. Then, one can find that $\lambda$ controls the variable step-size with the range of $\mu(n) \in (0, \mu]$. It is necessary to choose $\lambda$ properly to balance the algorithm stability and the steady-state MSE performance of LMS/F. For achieving the better steady-state MSE performance without scarifying algorithm stability, one may choose optimal $\lambda$ according to the method in [6]. Setting $e^2(n) = 0.1$ for example, $\lambda$ controls the variable step-size as shown in Fig. 3. If one fix the error term as $e^2(n)=0.1$, smaller $\lambda$ incurs smaller step-size $\mu(n)$ which ensures more stability and lower MSE but higher computational complexity (slower convergence speed) and vice versa.

To provide the insight of the behavior of LMS/F, its steady-state MSE is derived as follows. Let $\beta(\infty)$ denote the steady state ratio which is given by $\mu(\infty)/\mu$ with $\mu(\infty) = \lim_{n\to\infty} \mu(n)$.

**Theorem 1.** For the variable step-size $0 < \mu(n) < 2/(N+2)\sigma_x^2$, then steady state MSE of LMS/F is given by

$$D_{lmsf}(\infty)=\frac{\mu\beta(\infty)N\sigma_n^2}{2-\mu\beta(\infty)(N+2)\sigma_x^2}. \tag{5}$$

Proof: Denote $\phi(n)=\mu(n)/\mu$ as the ratio of VSS and ISS. From Eq. (4), one can get

$$\phi(n)=\frac{\mu(n)}{\mu}=\frac{e^2(n)}{e^2(n)+\lambda}. \tag{6}$$

Due to the unconditioned update of $e^2(n)$, the analysis of $\phi(n)$ is very difficult. Therefore, the conditional mean value $\beta(n)$ is introduced to approximate $\phi(n)$. Defining the n-th transient estimation error of channel estimator as $\mathbf{v}(n)=\tilde{\mathbf{h}}(n)-\mathbf{h}$, the conditional mean step-size ratio $\beta(n)$ can be derived as follows:

$$\begin{aligned}
\beta(n)&=E[\phi(n)|\mathbf{v}(n)]\\
&=\frac{1}{\sqrt{2\pi}\sigma_e(n)}\int_{-\infty}^{\infty}\frac{e^2(n)}{e^2(n)+\lambda}\exp\left(-\frac{e^2(n)}{2\sigma_e^2(n)}\right)de(n)\\
&=\frac{2}{\sqrt{2\pi}}\sqrt{\frac{\lambda}{\sigma_e^2(n)}}\int_0^{\infty}\left(1-\frac{1}{1+t^2(n)}\right)\exp\left(-\frac{\lambda t^2(n)}{2\sigma_e^2(n)}\right)dt(n)\\
&=1-\frac{2}{\sqrt{2\pi}}\sqrt{\frac{\lambda}{\sigma_e^2(n)}}\int_0^{\infty}\frac{1}{1+t^2(n)}\exp\left(-\frac{\lambda t^2(n)}{2\sigma_e^2(n)}\right)dt(n)\\
&=1-\sqrt{\frac{\pi\lambda}{2\sigma_e^2(n)}}\exp\left(\frac{\lambda}{2\sigma_e^2(n)}\right)\left[1-erf\left(\sqrt{\frac{\lambda}{2\sigma_e^2(n)}}\right)\right],
\end{aligned} \tag{7}$$

where $t(n)=e(n)/\sqrt{\lambda}$. From Eq. (8), one can find that conditional mean step-size $\beta(n)$ is a function of the ratio $\lambda/\sigma_e^2(n)$. $\sigma_e^2(n)$ is a conditional variance whose mean value is computed by

$$E[\sigma_e^2(n)]=E[e^2(n)|\mathbf{v}(n)]=\sigma_n^2+\sigma_x^2D_{lmsf}(n). \tag{8}$$

In [3], the steady-state MSE of LMS with step-size $\mu_s$ is derived as

$$D_{lms}(\infty)=\frac{\mu_sN\sigma_n^2}{2-\mu_s\sigma_x^2(N+2)}. \tag{9}$$

According to derivation in Eq. (8), LMS/F can be considered as LMS with a variable step-size $\mu\beta(n)$ [14]. Hence, the steady-state MSE of LMS/F with mean step-size $\mu\beta(n)$ is derived by

$$D_{lmsf}(\infty)=\frac{\mu(\infty)N\sigma_n^2}{2-\mu(\infty)\sigma_x^2(N+2)}.=\frac{\mu\beta(\infty)N\sigma_n^2}{2-\mu\beta(\infty)\sigma_x^2(N+2)}. \tag{10}$$

■

Theorem 1 states that $D_{lmsf}(\infty)$ depends on length $N$ of channel, i.e., longer channel length will incur higher MSE and vice visa. Based on the dense channel model assumption, $D_{lmsf}(\infty)$ in Eq. (11) is derived as the lower bound of linear channel estimator. However, the real channel is often modeled as sparse rather than dense. Hence, it is expected that the MSE performance can be improved by taking the advantage of channel sparsity. As a benchmark, the ideal steady-state MSE of LMS/F is also derived as follows.

**Theorem 2**. Assume that the position set $\Omega = \#\{l | h_l \neq 0, l = 0,1,...,N-1\}$ of nonzero channel taps is known. Suppose variable step-size $\mu(n)$ satisfies $0 < \mu(n) < 2/(N+2)\sigma_x^2$ to guarantee algorithm stability. Then oracle (ideal case) steady-state MSE $D_{orc}(\infty)$ [15] should satisfy

$$D_{orc}(\infty) = \frac{\mu\beta(\infty)K\sigma_n^2}{2 - \mu\beta(\infty)\sigma_x^2(N+2)}, \tag{11}$$

where $K$ is the number of nonzero channel taps. ■

Unfortunately, it is unable to know the exact position set before channel detection or estimation in practical communication systems. In the next section, two sparse LMS/F filtering algorithms are proposed to exploit channel sparsity so that they can improve MSE performance. In addition, to confirm the effectiveness, their corresponding theoretical performance bounds are also derived.

## 3. Improved sparse channel estimation using sparse LMS/F algorithms

### *3.1. Proposed ZA-LMS/F Algorithm*

Recall that the standard LMS/F in Eq. (2) does not make use of the channel sparsity. This is because the original cost function $G_{lmsf}(n)$ adopts neither sparse constraint nor penalty function. To exploit the channel sparsity, an sparse cost function is constructed by introducing an $\ell_1$-norm sparse constraint [11] to the cost function in Eq. (2). The proposed cost function is termed as zero-attracting LMS/F (ZA-LMS/F). The cost function of ZA-LMS/F is written as

$$G_{ZA}(n) = \underbrace{\frac{1}{2}e^2(n) - \frac{1}{2}\lambda \ln\left(e^2(n) + \lambda\right)}_{G_{lmsf}(n)} + \underbrace{\rho_{ZA}\left\|\tilde{\mathbf{h}}(n)\right\|_1}_{sparse\ constraint}, \tag{12}$$

where $\rho_{ZA}$ denotes a regularization parameter which balances the update error term $G_{lmsf}(n)$ and sparsity of the n-th channel estimator $\tilde{\mathbf{h}}(n)$.

For a better understanding of the difference between $G_{lmsf}(n)$ in Eq. (2) and $G_{ZA}(n)$ in Eq. (13), geometrical interpretation in two-dimensional signal plane $(x_1, x_2)$ is shown in Fig. 4. Hence, the constraints of $\ell_1$-norm and $\ell_2$-norm can be written as

$$f_{\ell 1}(x_1, x_2) = |x_1| + |x_2|, \tag{13}$$

$$f_{\ell 2}(x_1, x_2) = \sqrt{|x_1|^2 + |x_2|^2}. \tag{14}$$

In Fig. 4 shows that $f_{\ell 1}(x_1, x_2)$ has a square shape and $f_{\ell 2}(x_1, x_2)$ has a circle shape, respectively. The aim of norm constraint is to find unique solution (convex point) in the solution line which includes many potential solutions. It is well known that $\ell_2$-norm constraint can find many solutions but unable to decide the unique solution (convex point). Unlike the $\ell_2$-norm, $\ell_1$-norm constraint can find unique solution (convex point) [15]. In Fig. 4, one can find that $G_{lmsf}(n)$ cannot find sparse solutions (convex point) in the solution line. Different from $G_{lmsf}(n)$, the cost function $G_{ZA}(n)$ can find a unique sparse solution (convex point) in the solution line. From Eq. (13), the update equation of ZA-LMS/F is derived as

$$\begin{aligned}\tilde{\mathbf{h}}(n+1) &= \tilde{\mathbf{h}}(n) + \mu \frac{\partial G_{ZA}(n)}{\partial \tilde{\mathbf{h}}(n)} \\ &= \tilde{\mathbf{h}}(n) + \frac{\mu e^3(n)\mathbf{x}}{e^2(n)+\lambda} - \gamma_{ZA}\,\mathrm{sgn}\left\{\tilde{\mathbf{h}}(n)\right\}, \\ &= \tilde{\mathbf{h}}(n) + \mu(n)e(n)\mathbf{x} - \gamma_{ZA}\,\mathrm{sgn}\left\{\tilde{\mathbf{h}}(n)\right\},\end{aligned} \tag{15}$$

where $\gamma_{ZA} = \mu\rho_{ZA}$ is a positive parameter. Based on above update equation, the convergence property of ZA-LMS/F is derived in Theorem 3.

**Theorem 3**. Suppose variable step-size $\mu(n)$ satisfies $0 < \mu(n) < 2/(N+2)\sigma_x^2$ . As $n \to \infty$, the mean channel vector $E\{\tilde{\mathbf{h}}(n)\}$ obtained by ZA-LMS/F converges to

$$E\{\tilde{\mathbf{h}}(\infty)\} = \mathbf{h} - \frac{\gamma_{ZA}}{\mu\beta(\infty)}\mathbf{R}^{-1}E\left\{\mathrm{sgn}\left(\tilde{\mathbf{h}}(\infty)\right)\right\}. \tag{16}$$

Proof. By subtracting channel vector $\mathbf{h}$ from both sides of Eq. (14), one obtain

$$\begin{aligned}\mathbf{v}(n+1) &= \mathbf{v}(n)+\mu(n)e(n)\mathbf{x}-\gamma_{ZA}\operatorname{sgn}\left(\tilde{\mathbf{h}}(n)\right)\\ &= \mathbf{v}(n)+\mu(n)\left(z-\mathbf{v}^T(n)\mathbf{x}\right)\mathbf{x}-\gamma_{ZA}\operatorname{sgn}\left(\tilde{\mathbf{h}}(n)\right)\\ &= \left(\mathbf{I}-\mu(n)\mathbf{x}\mathbf{x}^T\right)\mathbf{v}(n)+\mu(n)z\mathbf{x}-\gamma_{ZA}\operatorname{sgn}\left(\tilde{\mathbf{h}}(n)\right).\end{aligned} \tag{17}$$

where $\mathbf{v}(n)=\tilde{\mathbf{h}}(n)-\mathbf{h}$ denotes the n-th adaptive channel estimation error. Taking expectations on both sides of (18) and setting $n\rightarrow\infty$, one can obtain

$$E\left\{\mathbf{v}(\infty)\right\}=\left(\mathbf{I}-\mu\beta(\infty)\mathbf{R}\right)E\left\{\mathbf{v}(\infty)\right\}-\gamma_{ZA}E\left\{\operatorname{sgn}(\tilde{\mathbf{h}}(\infty))\right\}. \tag{18}$$

Please notice that $E[\mu(n)|\mathbf{v}(n)]=\mu\beta(n)$ was considered in Eq. (18) according to Eqs. (8) and (9). Thus, the proof of Theorem 3 is arrived. ■

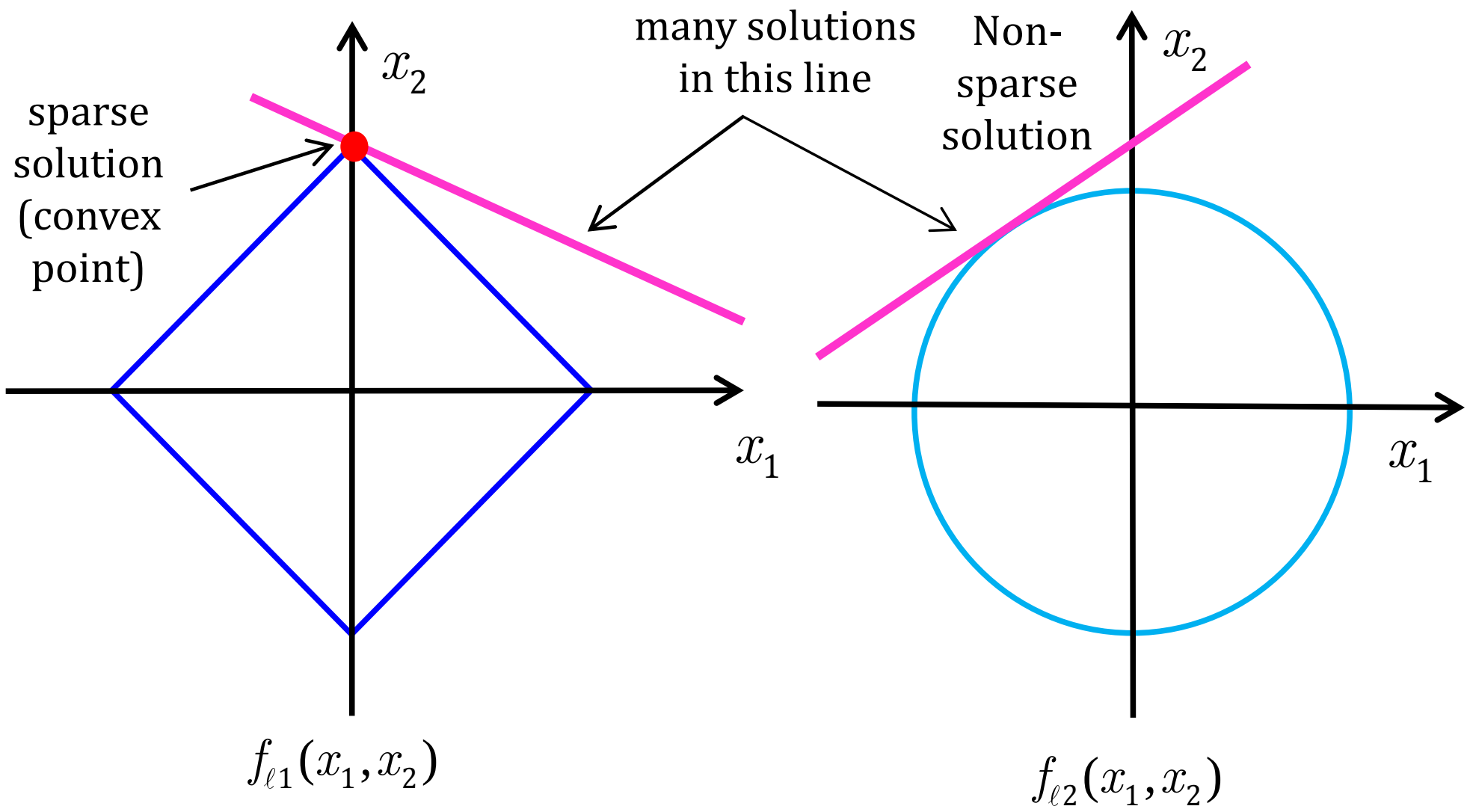

Fig. 4. Geometrical interpretation of sparse solution is obtained by using sparse $\ell_1$-norm sparse constraint function.

According to Eq. (13), ZA-LMS/F is equivalent to ZA-LMS with variable step-size $\mu(n)$. Hence, steady state MSE bound of ZA-LMS/F is implied in [15] as

$$\begin{aligned}D_{ZA}(\infty) = D_{lmsf}(\infty) &- \frac{\gamma_{ZA}(N-K)\sqrt{p}}{\sqrt{2\pi}\mu^2\beta^2(\infty)\sigma_x^4\Delta_N^2} + \frac{\gamma_{ZA}^2\left(2(N-K)\right)\Delta_\mu\Delta_K}{\pi\mu^2\beta^2(\infty)\sigma_x^4\Delta_N^2}\\ &+\frac{\pi\Delta_N(\mu\beta(\infty)\sigma_x^2+2K\Delta_\mu)}{\pi\mu^2\beta^2(\infty)\sigma_x^4\Delta_N^2},\end{aligned} \tag{19}$$

where $\Delta_\mu = 1 - \mu\beta(\infty)$ and $\Delta_K = 2 - (K+2)\mu\beta(\infty)\sigma_x^2$, and $p$ is a discriminant which is given by

$$p = 8\gamma_{ZA}^2 \Delta_K^2 \Delta_\mu^2 / \pi + 16\mu\beta(\infty)\sigma_x^2 \Delta_N \Delta_\mu^2 \left( \gamma_{ZA}^2 (K+1) + \mu^2 \beta^2(\infty) \sigma_x^2 \sigma_n^2 \right). \tag{20}$$

Note that the equivalent result in Eq. (20) was also found in [16]. According to the bound, one can find that selecting proper $\gamma_{ZA}$ for ZA-LMS/F can achieve lower MSE bound than standard LMS/F.

### *3.2. Proposed RZA-LMS/F Algorithm*

The ZA-LMS/F cannot distinguish zero taps and non-zero taps effectively due to the fact all of the taps are forced to zero uniformly as show in Fig. 5. Motivated by reweighted $\ell_1$-minimization (RL1) sparse recovery algorithm [12] in compressed sensing (CS) [17], [18], an improved ASCE method is proposed with RZA-LMS/F algorithm. The cost function of RZA-LMS/F is constructed by

$$G_{RZA}(n) = \underbrace{\frac{1}{2}e^2(n) - \frac{1}{2}\lambda \ln\left(e^2(n) + \lambda\right)}_{G_{lmsf}(n)} + \underbrace{\rho_{RZA} \sum_{i=0}^{N-1} \log(1 + \varepsilon h_i(\mathrm{n}))}_{sparse\ constraint}, \tag{21}$$

where $\rho_{RZA} > 0$ is a sparse regularization parameter which balances the estimation error and channel sparsity; $\varepsilon$ is the reweighted factor to exploit much more channel sparsity than ZA-LMS/F. The intuitive understanding of reweighted $\ell_1$-minimization sparse recovery algorithm is that the estimated channel taps with small value are forced to zero. Based on the cost function in Eq. (20), the corresponding update equation is

$$\begin{aligned} \tilde{\mathbf{h}}(n+1) &= \tilde{\mathbf{h}}(n) + \mu \frac{\partial G_{RZA}(n)}{\partial \tilde{\mathbf{h}}(n)} \\ &= \tilde{\mathbf{h}}(n) + \frac{\mu e^3(n)\mathbf{x}}{e^2(n) + \lambda} - \gamma_{RZA} \frac{\mathrm{sgn}(\tilde{\mathbf{h}}(n))}{1 + \varepsilon \left|\tilde{\mathbf{h}}(n)\right|}, \end{aligned} \tag{22}$$

where $\gamma_{RZA} = \varepsilon\mu\rho_{RZA}$ is a parameter which depends on three parameters: $\mu$, $\rho_{RZA}$ and $\varepsilon$.

One can easy find that parameter selection is very important and crucial for achieving reasonable tradeoff between steady-state MSE performance, convergence speed and sparsity exploiting. In addition, the estimated channel taps $\left|\tilde{h}_i(n)\right|$, $i = 0, 1, \ldots, N-1$ are replaced by zeroes in high probability if these taps are smaller than given threshold $1/\varepsilon$ in the second term of Eq. (23). Hence, reweighted factor $\varepsilon$ is very important for RZA-LMS/F. That is to say, RZA-LMS/F utilizes proper reweighted factor can achieve better performance than ZA-LMS/F while improper one may cause worse performance than ZA-LMS/F.

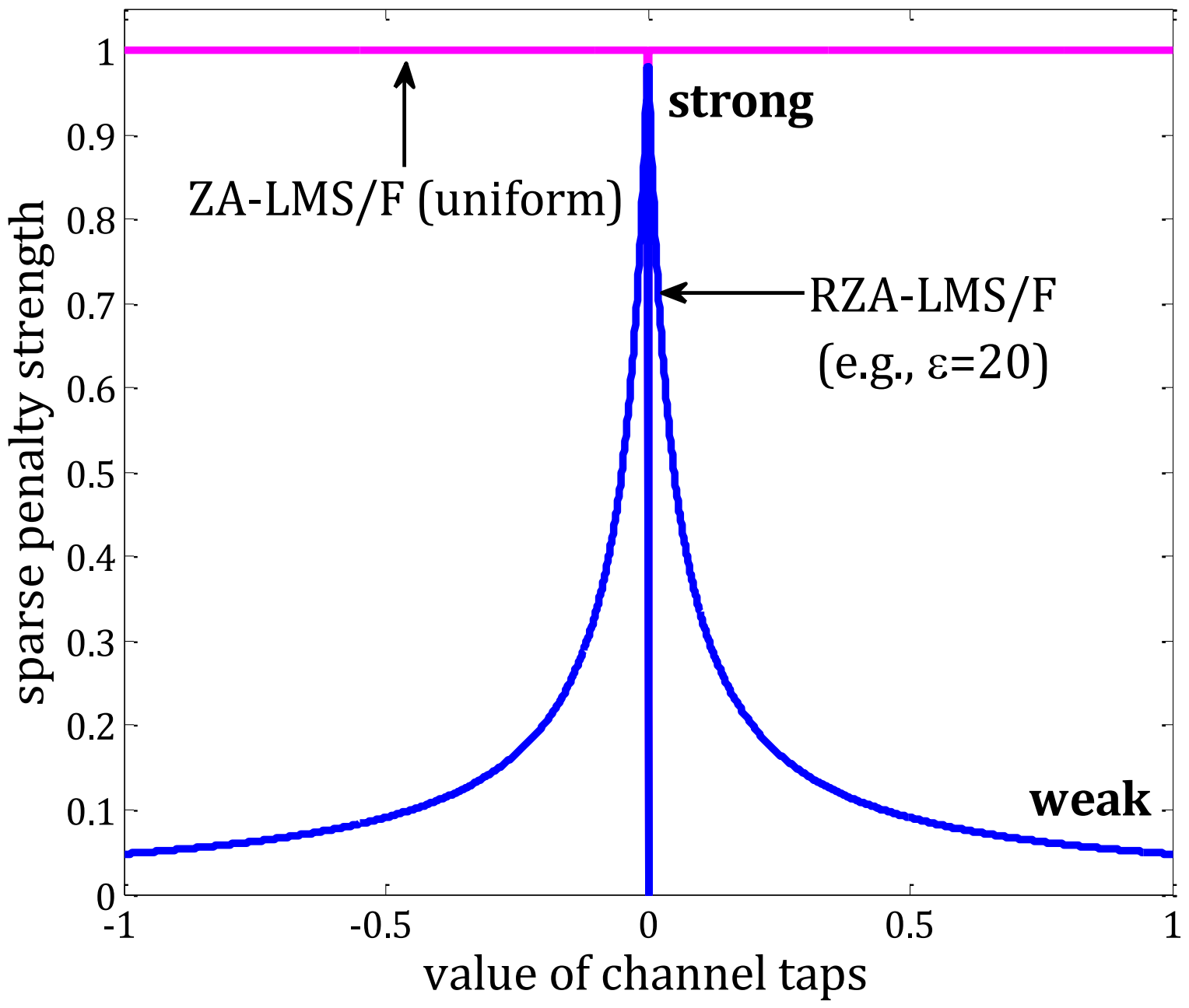

Fig. 5. Sparse penalty strength over different value of channel taps.

### *3.3. Regularization parameter Selection of sparse LMS/F filtering algorithms*

It is well known that regularization parameter is very important for LASSO based sparse channel estimation [11]. In [19], a parameter selection method was proposed for LASSO based partial sparse channel estimation. To the best of authors' knowledge, however, there is no report on regularization parameter selection method for sparse LMS/F algorithms, i.e., ZA-LMS/F and RZA-LMS/F. It is worth noticing that the optimal regularization parameter selection is an NP hard problem. For one thing, optimal sparse solution of sparse LMS/F algorithms should exploit all of position information of nonzero taps but it is impossible to estimation sparse channels in noisy environment. For another, regularization parameter could be selected adaptively by learning the channel structure and noise level. But it requires extra high computational cost which is also one of important system evaluation criteria. To avoid the above mentioned problems, empirical regularization parameter is selected for sparse LMS/F algorithms via Monte Carlo methods. Empirical selection method is formulated as follows.

Computer simulation adopts 10000 independent runs for achieving average MSE performance. Simulation parameters are given in Tab. 1. The estimation performance is evaluated by average steady-state MSE metric which is defined as

$$\text{Average MSE}\{\tilde{\mathbf{h}}(n)\} = E\left\{\|\mathbf{h}-\tilde{\mathbf{h}}(n)\|_2^2\right\}. \tag{23}$$

Utilizing different regularization parameters, MSE performance curves of ZA-LMS/F and RZA-LMS/F are depicted in Figs. 6 and 7, respectively. Fig. 6 shows that MSE performance is near optimal with $\rho_{ZA}=0.0004$ and $\rho_{ZA}=0.0002$ produce smaller average MSE performance for $K=2$ and $K=4$, respectively. Likewise, as shown in Fig. 7, choosing approximate optimal regularization parameters $\rho_{RZA}=0.06$ and $\rho_{RZA}=0.04$ for RZA-LMS/F can achieve smaller average MSE performance for $K=2$ and $K=4$, respectively. In the following, these parameters will be utilized for performance comparison with sparse LMS algorithms in Section 4.

Table 1. Simulation parameters

| **Parameters** | | **values** |
|---|---|---|
| Training signal | | Pseudorandom noise (PN) sequence |
| Fading channel | Fading | Frequency-selective fading |
| | Length | $N$ =16 |
| | Nonzero taps | $K$ =2 and 4 |
| | Taps' distribution | Random Gaussian $\mathcal{CN}(0,1)$ |
| Threshold parameter for LMS/F-type algorithms | | $\lambda = 0.8$ |
| Average SNR | | 10dB |
| Initial step-size | | $\mu = 0.04$ |
| Re-weighted factor for RZA-LMS/F | | $\varepsilon = 20$ |

Remark: One can find that the regularization parameter depends upon the number of nonzero taps of channels. Smaller regularization parameter should be selected for sparser channel, vice versa. On real channel estimation, even though the number of nonzero taps is unknown, empirical regularization parameter can still selected in some range while the MSE performance gaps are not obvious in the case of different number of nonzero taps as shown in Figs. 6 and 7. In this paper, to compare the performance of the proposed algorithms fairly, both sparse LMS algorithms and sparse LMS/F ones are adopted approximated optimal empirical regularization parameters. Therefore, in this paper, one can assume that $K$ is priori known.

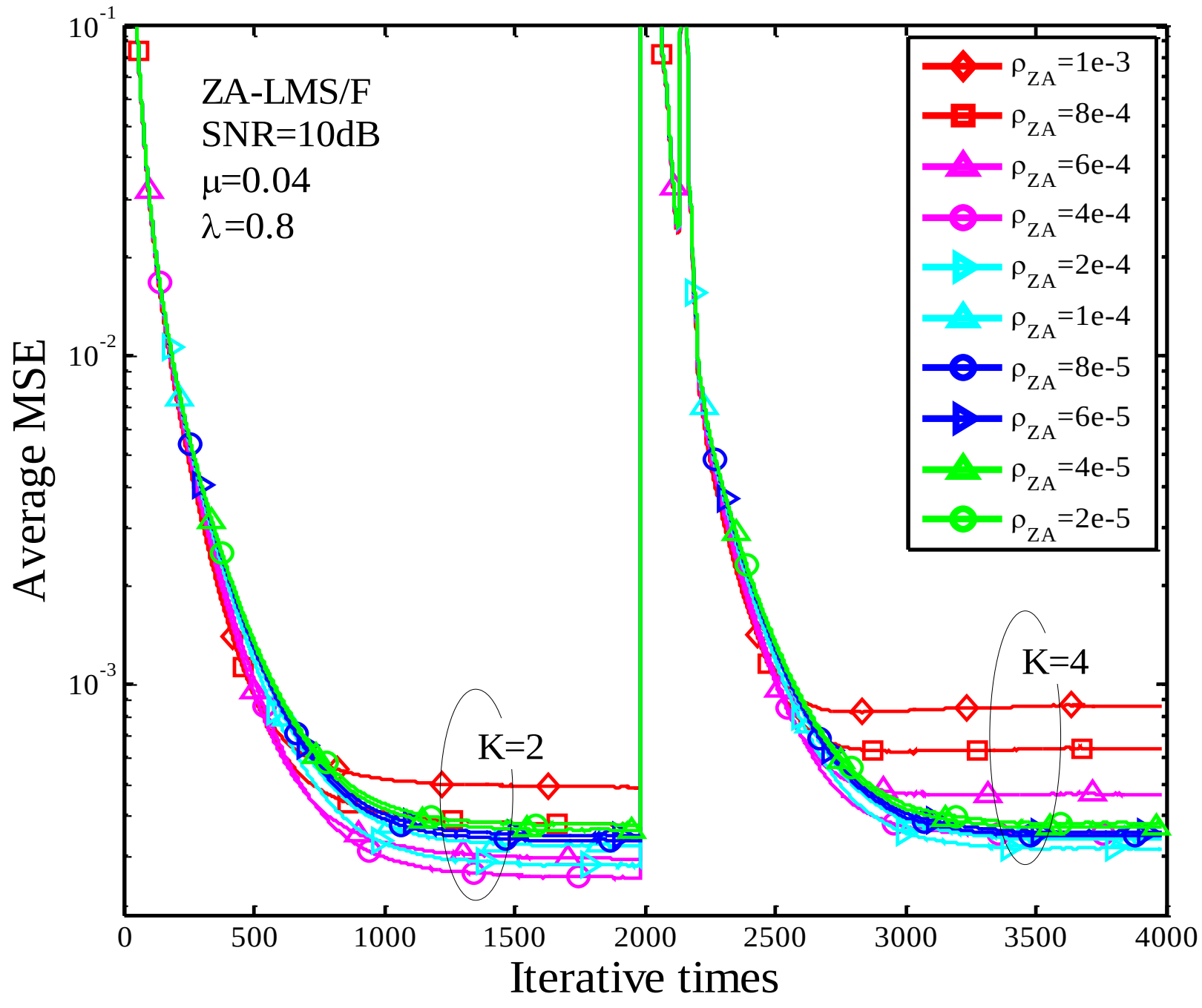

Fig. 6. The impact of regularization parameter $\rho_{ZA}$ on average MSE.

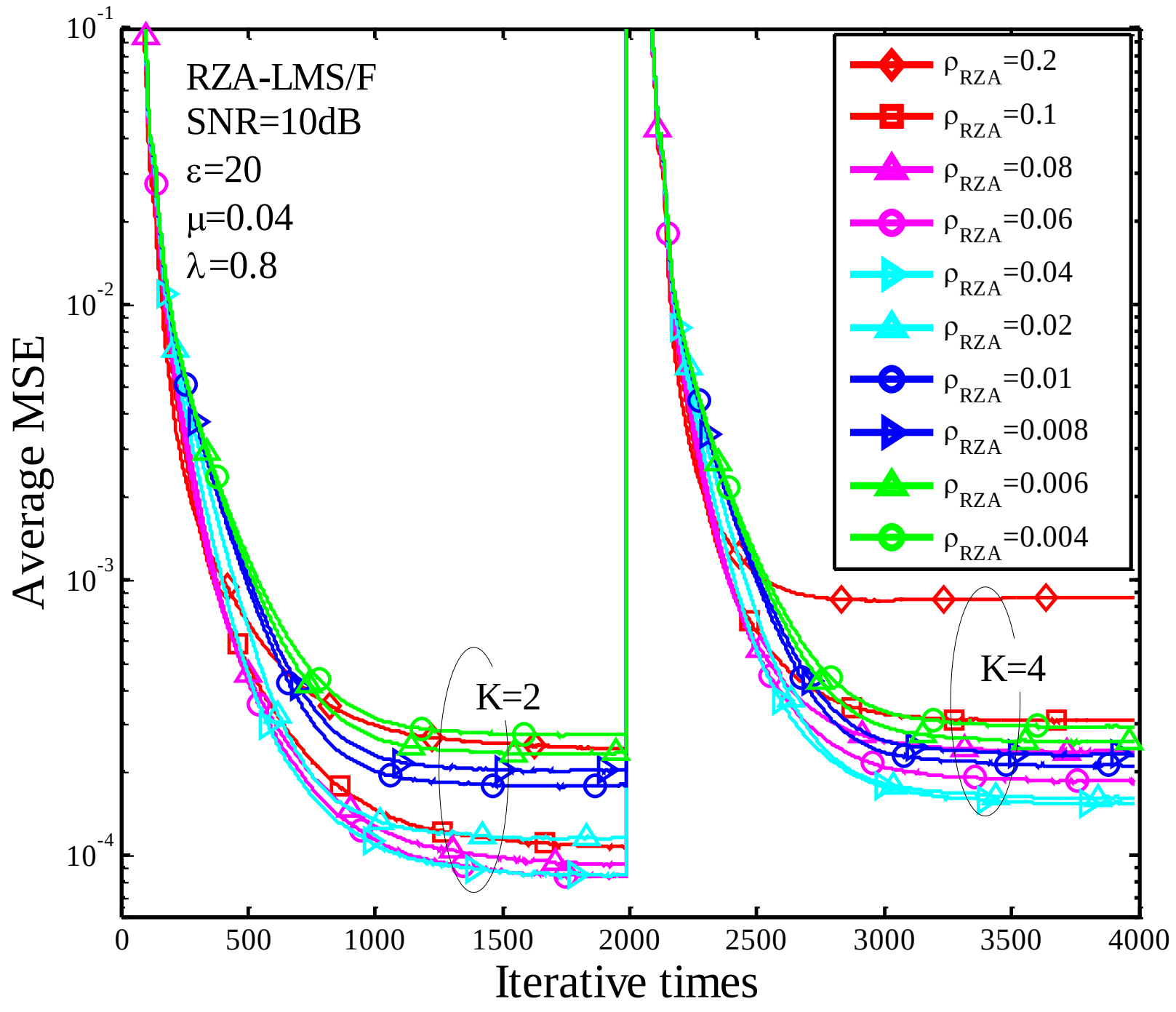

Fig. 7. The impact of regularization parameter $\rho_{RZA}$ on average MSE.

## 4. Computer Simulations

In this section, the average MSE performance of the proposed ASCE methods using (R)ZA-LMS/F algorithms is evaluated. The results are averaged over 1000 independent Monte-Carlo runs. The length of channel vector $\mathbf{h}$ is set as $N = 16$ and its number of dominant taps is set to $K = 2$ and 4, respectively. Each dominant channel tap follows random Gaussian distribution as $\mathcal{CN}(0, \sigma_{\mathbf{h}}^2)$ which is subject to $\mathrm{E}\{||\mathbf{h}||_2^2 = 1\}$ and their positions are randomly decided within the length of $\mathbf{h}$. The received signal-to-noise ratio (SNR) is defined as $10\log(E_s/\sigma_n^2)$, where $E_s = 1$ is the unit transmission power. Here, the SNR is set as 10dB. All of the step-sizes and regularization parameters are listed in Tab. 2.

Firstly, average MSE performance of proposed methods is evaluated for $K$=2 and 4. To confirm the effectiveness of the proposed methods, they are compared with sparse LMS algorithms, i.e., ZA-LMS and RZA-LMS [20].

Table 2. Simulation parameters

| Parameters | | values |
|---|---|---|
| Training signal | | PN sequence |
| Fading channel | Fading | Frequency-selective fading |
| | Length | $N$=16 |
| | Nonzero taps | $K$=2 and 4 |
| | Distribution | Random Gaussian $\mathcal{CN}(0,1)$ |
| Threshold parameter for LMS/F-type algorithms | | $\lambda = 0.8$ |
| Average SNR | | 10dB |
| Step-size for gradient descend | | $\mu = 0.04$ (for LMS/F-type) |
| | | $\mu_S = 0.008$ (for LMS-type) |
| Regularization parameter for $K$=2 | | $\rho_{ZA} = 0.0004$ and $\rho_{RZA} = 0.06$<br>$\rho_{\mathrm{ZAS}} = 0.008$ and $\rho_{RZAS} = 0.8$ |
| Regularization parameter for $K$=4 | | $\rho_{ZA} = 0.0002$ and $\rho_{RZA} = 0.04$<br>$\rho_{ZAS} = 0.004$ and $\rho_{RZAS} = 0.4$ |
| Re-weighted factor of RZA-LMS(/F) | | $\varepsilon = 20$ |

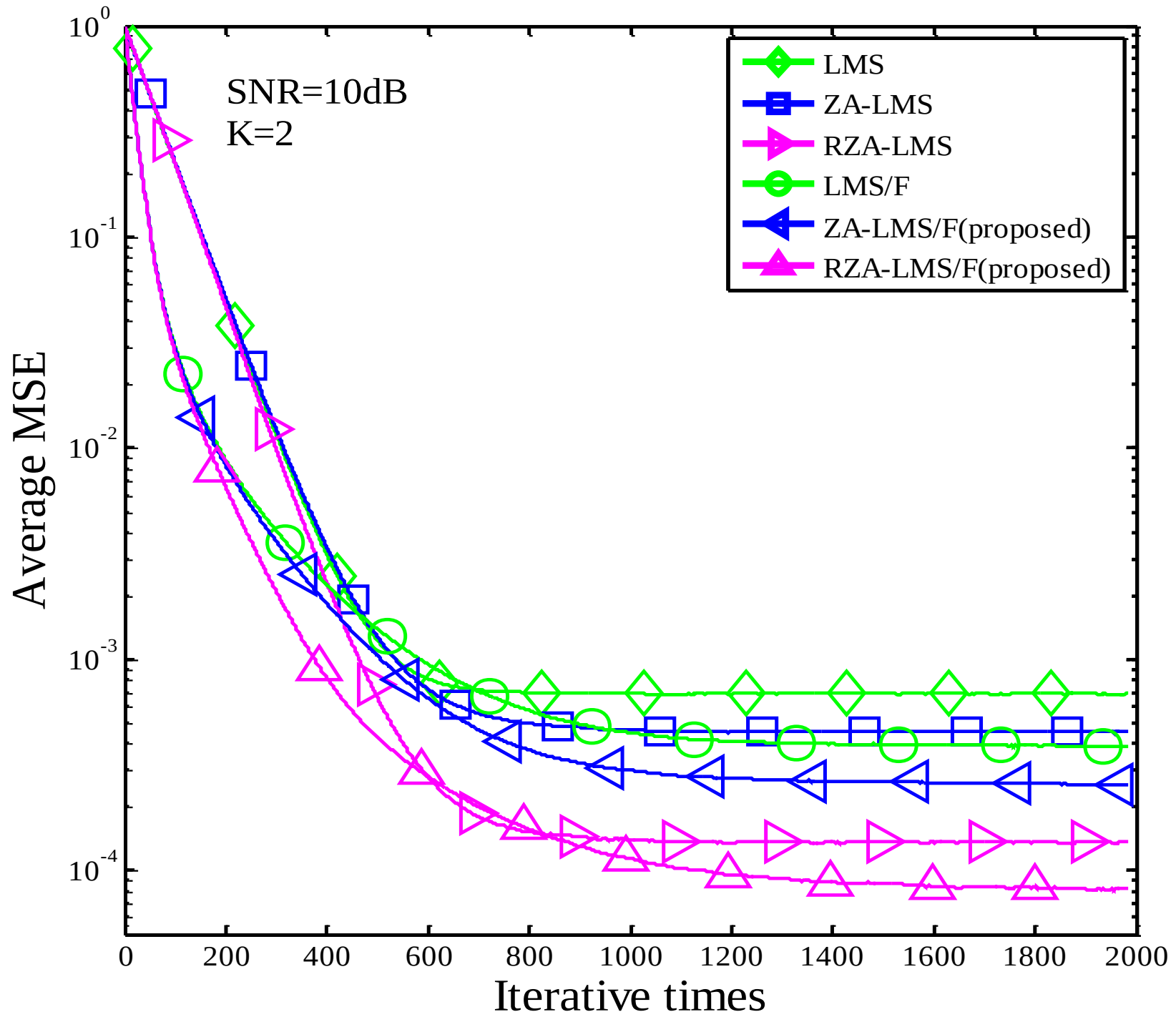

Fig. 8. Performance comparisons with respect to $K$=2.

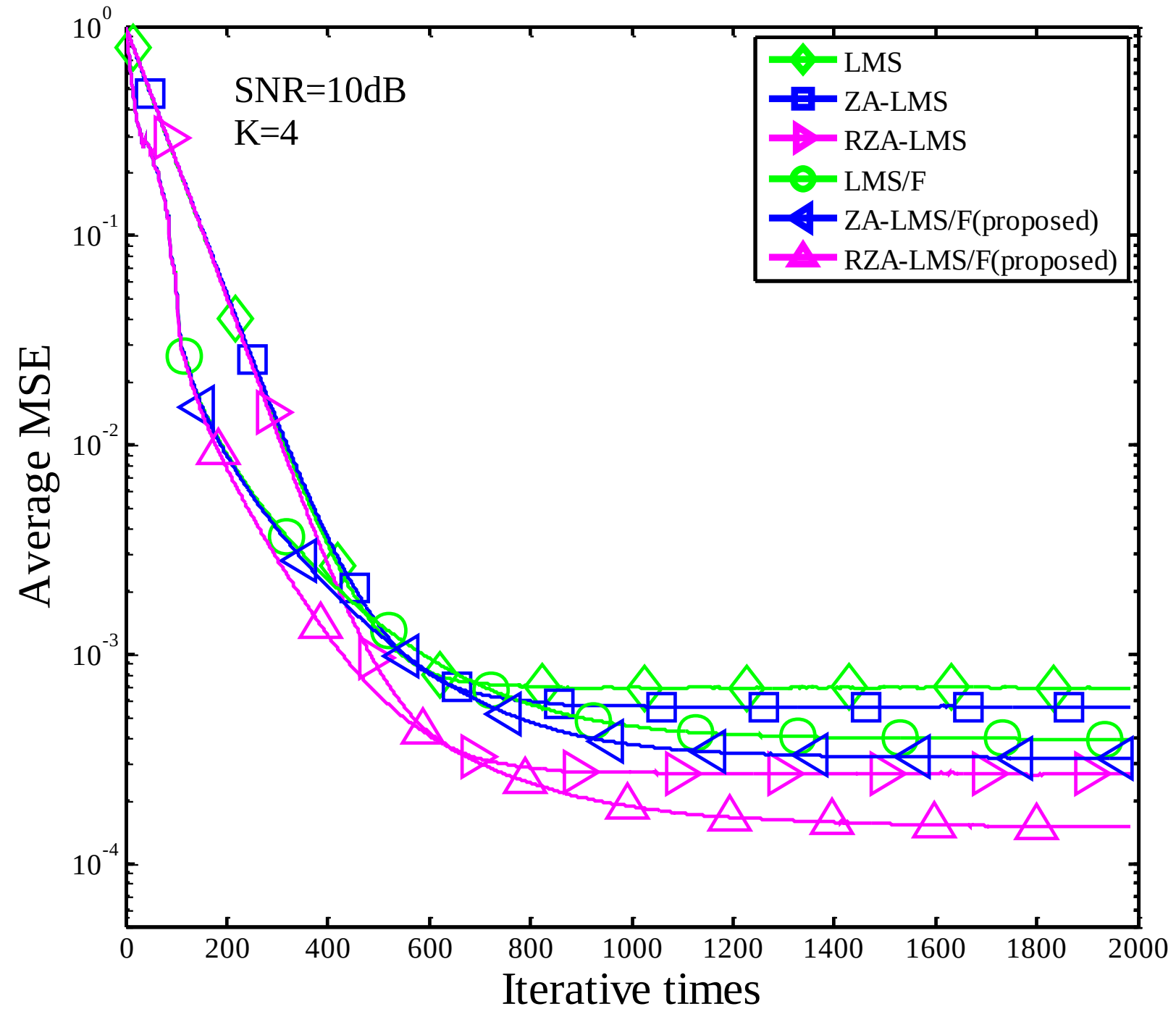

Fig. 9. Performance comparisons with respect to $K$=4.

It is well known that the step-size $\mu_s$ of LMS-type algorithms is invariable during the process of gradient descend. However, the step-size $\mu$ of LMS/F algorithms is variable since it depends on instantaneous estimation error. For a fair comparison, $\mu_s = 0.008$ and $\mu = 0.04$ are selected for LMS-type algorithms and LMS/F ones, respectively. In addition, to achieve better steady-state estimation performance, regularization parameters for two sparse LMS algorithms are adopted from the paper [21], i.e., $\rho_{ZAS} = 0.008$ and $\rho_{RZAS} = 0.8$ for $K = 2$; $\rho_{ZAS} = 0.004$ and $\rho_{RZAS} = 0.4$ for $K$ =4. Average MSE performance comparison curves are depicted in Fig. 8 and Fig. 9, respectively. LMS/F algorithms achieve better estimation performance than LMS algorithms in [12]. Since $\mu > \mu_s$, the convergence speed of proposed algorithms is faster than LMS-type algorithms. In addition, figures clarify that the sparse LMS/F algorithms, i.e., ZA-LMS/F and RZA-LMS/F, achieve better estimation performance than LMS/F due to the fact that sparse LMS/F algorithms utilize $\ell_1$-norm sparse constraint function.

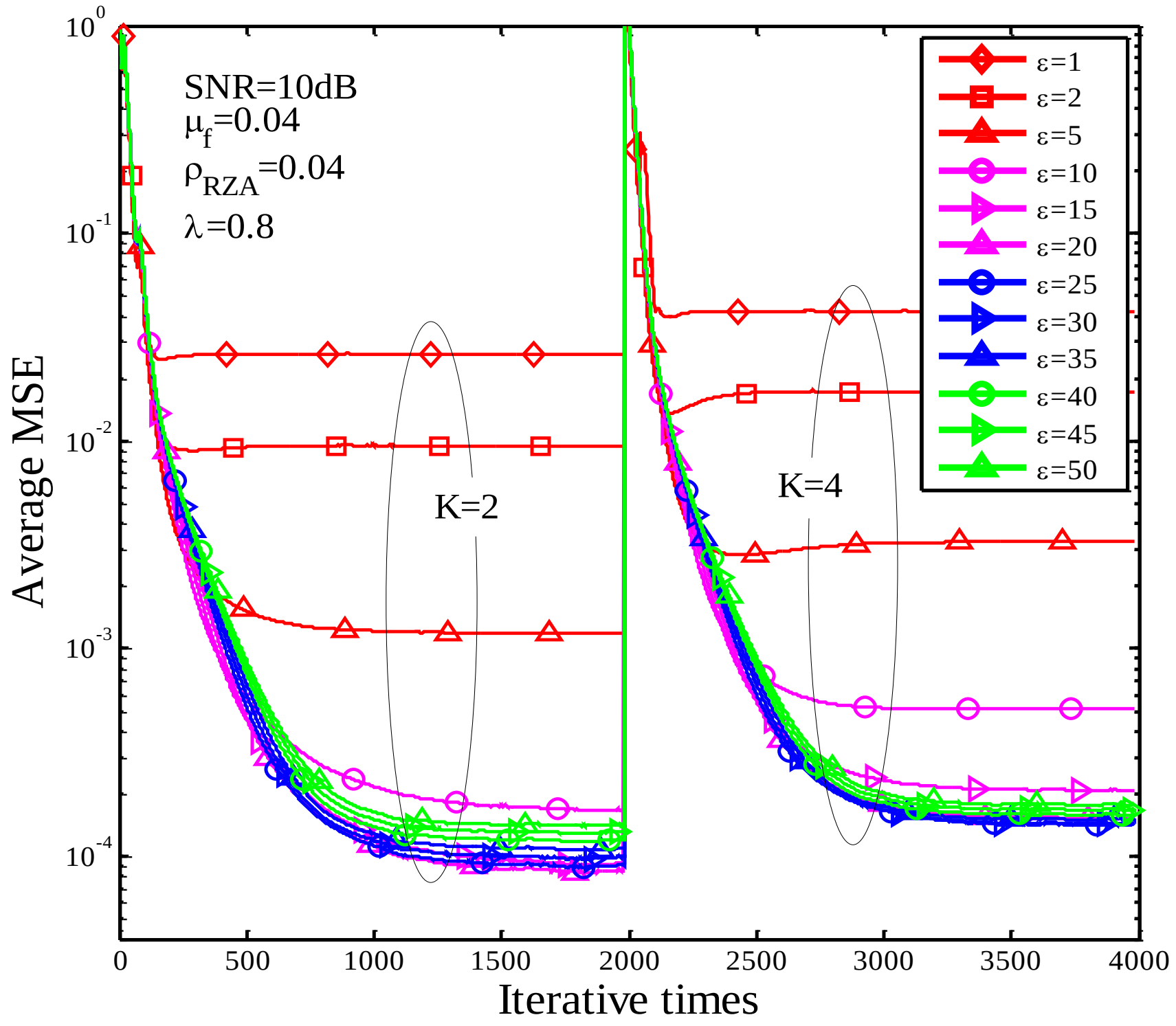

Fig. 10. MSE performance of RZA-LMS/F v.s. re-weighted factor $\varepsilon$.

Secondly, MSE performance curves of RZA-LMS/F with different reweighted factors $\varepsilon \in \{1,2,5,\dots,50\}$ are shown in Fig. 10 for $K=2$ and 4, respectively. Under the simulation setup considered, RZA-LMS/F using $\varepsilon=20$ or 25 can achieve near optimal estimation performance. Fig. 10 shows that the performance of the RZA-LMS/F depends highly on reweighted factor $\varepsilon$. Since impropriate ε may degrade the estimation performance. Hence, the proper selection of the reweighted factor is also important for the RZA-LMS/F on ASCE.

## 5. Conclusions and future work

In this paper, improved SFEC-ASCE method using sparse LMS/F algorithms were proposed for estimating channels in sparse multipath environments. Based on the CS theory, an SFEC-ASCE method with ZA-LMS/F algorithm was first proposed to exploit channel sparsity so that it can improve the estimation performance when comparing to traditional LMS/F. Corresponding steady-state MSE performance bound was derived to demonstrate the performance advantage as well. Furthermore, inspired by re-weighted $\ell_1$-norm algorithm in CS, an second improved SFEC-ASCE method using RZA-LMS/F algorithm was also proposed to exploit much more sparse information than former proposed method. By virtual of Monte Carlo simulation, a simple method is proposed for choosing the approximate optimal regularization parameter of sparse LMS/F algorithms. Simulation results showed that the proposed algorithms achieved better MSE performance than any sparse LMS filtering algorithms and traditional LMS/F filtering algorithms.

Since the optimal regularization parameter of sparse LMS/F algorithms is very important to balance the estimation error and channel sparsity. Only the Monte Carlo based approximate parameter selection method was considered in this paper. In future work, one can try to find the optimal regularization parameter in numerical analysis. It is expected the optimal parameter changed adaptively with respect to channel sparseness and estimation error.

## Acknowledgment

The authors would like to think Dr. Koichi Adachi of Institute for Infocomm Research for his valuable comments and helpful suggestions. This work was supported in part by the Japan Society for the Promotion of Science (JSPS) research activity start-up research grant (No. 26889050), Akita Prefectural University startup research grant as well as the National Natural Science Foundation of China grants (No. 61401069, No. 61261048, No. 61201273).